\documentclass[10pt,conference]{IEEEtran}

\usepackage{amssymb}
\usepackage{amsxtra}
\usepackage{amsmath}
\usepackage{latexsym}
\usepackage{graphics}
\usepackage{verbatim}
\usepackage{epsfig}
\usepackage{setspace}
\usepackage{theorem}
\usepackage{algorithmic}
\usepackage{algorithm}

\theorembodyfont{\normalfont\slshape}

\newtheorem{theorem}{Theorem}[section]
\newtheorem{lemma}[theorem]{Lemma}
\newtheorem{proposition}[theorem]{Proposition}
\newtheorem{corollary}[theorem]{Corollary}

\newtheorem{remark}[theorem]{Remark}

\def\N{{\mathbb N}}

\def\v{{\mathbf v}}

\def\cC{{\mathcal C}}
\def\cD{{\mathcal D}}

\def\cL{{\mathcal L}}

\def\cP{{\mathcal P}}

\def\cW{{\mathcal W}}

\begin{document}

\thispagestyle{empty}

\title{A Graph Theoretical Approach for Network Coding\\
in Wireless Body Area Networks}

\author{\authorblockN{Eimear Byrne, Akiko Manada}
\authorblockA{School of Mathematical Sciences\\
University College Dublin\\
Belfield, Dublin 4, Dublin, IRELAND\\
\texttt{ebyrne@ucd.ie, akiko.manada@ucd.ie}}
\and \authorblockN{Stevan Marinkovic and Emanuel Popovici}
\authorblockA{Department of Microelectronic Engineering\\
University College Cork, UCC\\
Cork, IRELAND\\
\texttt{stevanm@ue.ucc.ie, e.popovici@ucc.ie }}
}
\date{}
\maketitle

\renewcommand{\thefootnote}{$*$}
\footnotetext{This work was supported in part by a  Science Finance of Ireland (SFI) of Ireland Grant 06/MI/006 and Science Foundation Ireland Grant Number 07/SRC/I1169.}

\renewcommand{\thefootnote}{\arabic{footnote}}
\setcounter{footnote}{0}

\begin{abstract} 
Modern medical wireless systems, such as 
wireless body area networks (WBANs), are applications of 
wireless networks that can be used as a tool of data transmission between 
patients and doctors. Accuracy of data transmission is an important requirement 
for such systems. % in order to enhance the secure and proper cares for patients. 
In this paper, we will propose 
a WBAN which is robust against erasures and describe its properties
using graph theoretic techniques. 
%More precisely, we will provide some properties for a WBAN to be 
%strong against the erasure of packets using graph theoretical techniques. 
\end{abstract}

%%%%%%%%%%%%%%%%%%%%%%%%%%%%%%%%%%%%%%%%%%%%%%%%%%%%%%%%%%%%%%

\section{Introduction\label{intro}}
Network coding has been widely studied in the last decade since the publication of the seminal paper \cite{A}
in which it was shown that significant gains could be achieved in a multicast transmission 
if coding of data is used in addition to simply routing.
One of the applications of network coding is 
to \emph{wireless body area networks} (WBANs) \cite{CGVCL},
which could offer valuable support to monitoring a person's physiological data.
Such systems are now more practical with the advent of new generation miniature, low-power wireless devices. 
A WBAN is a network that sends a person's relevant health information from attached or implanted miniature sensors, via relays, to a monitoring station (MS).
Such systems allow continuous remote updating, which has the potential to offer many advantages 
in modern medical care, allowing greater patient freedom and improved response to acute situations.

An important consideration of WBAN design is that the MS can retrieve 
all information sent in spite of packet loss.
At the same time, power-consumption and communication range of sensors should be also taken into consideration, since these sensors must be small and light, with small batteries and antennae. 
An efficient WBAN must be operable under very low transmission power compared with general wireless networks,
so any coding scheme for a given WBAN should have low computational complexity.
Another requirement is to minimise the number of re-transmissions requested due to errors.

In \cite{MP} a simple WBAN coding scheme robust to packet erasures was presented.
In this paper, we generalize that scheme, taking a graph theoretic perspective.
More precisely, we consider a graph which 
represents a coding scheme for a WBAN, and 
use it to analyze the robustness of the scheme against packet loss. 

We present preliminaries in 
Section~\ref{background}, and describe a graph representation of a WBAN coding scheme.
In Section~\ref{decodable} we give
a necessary and sufficient condition for a given WBAN coding scheme to 
be able to retrieve all data at its MS after the erasure of some packets, 
in terms of its corresponding graph representation, in which case we call the graph decodable.
%together with the definition of decodable graphs.
In Section~\ref{decodable_probability}, we give an expression for the {\em decoding probability} 
of the given WBAN scheme and provide a partial characterization of those with a high decoding probability.
%We also establish an algorithm to construct such a WBAN. 
In Section~\ref{simulation}, we present simulation data for a given WBAN.

\section{Basic Background\label{background}}
%\subsection{Preliminaries}
We begin with some background on WBANs (cf. \cite{CGVCL}, \cite{MP}).
A WBAN consists of \emph{sensors} $S_i$, \emph{relays} $R_j$ and a \emph{monitoring station} (MS). 
These sensors might be implanted, attached to
a person's skin or clothing or in the proximity of the body.
Each sensor $S_i$ sends a packet $P_i$ (a vector over $GF(2)$) to one or more relays where packets are encoded by taking linear combinations of them. The relays then send the encoded packets to the MS.
A {\em coding scheme} for a given WBAN is a collection of $GF(2)$-linear vectorial functions $f_{a}(P_1,...,P_n)$, corresponding to packet encodings at each relay. 

We consider schemes with \emph{redundancy} $r$, in which case
each packet is sent to $r$ different relays. 
Now let $n$ and $k$ be the number of sensors and relays, respectively, 
and assume that  each relay receives and sends $t$ packets. 
Then for a WBAN with redundancy $r$, observe that $rn$, the total number of packets sent from sensors to relays,  
is equal to $tk$, the total number of packets sent from relays to the MS. 
For the sake of simplicity, we assume the following throughout this paper.
\begin{itemize}
\item  $t=sr$ for some $s \in \N$ and therefore, $n=sk$ holds. Since $n>k$ in general, we have
$s\geq 2$.
\item Relay $R_j$ receives and encodes $t$ packets $P_{js+1}, P_{js+2}, \ldots, P_{js+t}$ where $js+\ell$ is computed modulo $n$ for each $j=0,1,\ldots, k-1$.   
\item Each encoded outgoing packet (to be sent to the MS) has the form $P_i$ or $P_i\oplus P_{i'}$, $i\not=i'$,
where as usual $\oplus $ denotes addition of binary vectors. In other words, $f_a(P_1,...,P_n)$ is either $P_i$ or $P_i\oplus P_{i'}$ 
for some $i,i'$. 
\item Erasures do not occur in communication from sensors to relays but some may occur in the communication from relays to the MS. 
\end{itemize}

We next present some preliminaries on graphs, (see \cite{W} for further reading).
Let $G$ be a finite graph with vertex set $V(G)$ and edge set $E(G)$, respectively.
We write $L_G$ to denote the number of loops of $G$. 
We define the \emph{incidence degree} of a vertex $v$, expressed $d_{I}(v)$, as  
the number of edges incident with $v$ (each loop at $v$ contributes a count of one to this number; of course this is different to the standard definition of the degree of $v$, 
in which loops contribute a count of two to the degree). 
We denote by $\delta _{I}(G)$ the minimum incidence degree of $G$, that is, $\delta _{I}(G)=\min \{d_{I}(v): v \in V(G)\}$. We write $\delta_\ell(v)$ to denote the number of loops incident with a given vertex $v$, and we let $\Delta_\ell(G):=\max\{ \delta_\ell(v) : v \in V(G)\}$.
For any graph $G$, it is well known that 
the sum of degrees of vertices in $G$ is equal to $2|E(G)|$. On the other hand, the sum of the incidence degrees $S_I(G)$ is given by $S_I(G)=2|E(G)|-L_G$
since each loop is counted as one edge.

A graph $G$ is called \emph{connected} if there is a path connecting each pair of vertices, otherwise $G$ is called \emph{disconnected}. 
Given a connected graph $G$, the \emph{edge-connectivity} $\kappa _G$ of $G$ is the smallest number of edges such that the resulting graph formed by deleting those edges is disconnected. 
Observe that $\kappa _G\leq \delta _I(G)$ since deleting all edges attached to a vertex $v$ with incidence degree $d_{I}(v)= \delta _{I}(G)$
makes $v$  isolated. 
A \emph{subgraph} $H$ of a graph $G$ is a graph such that $V(H) \subset V(G)$ and 
$E(H) \subset E(G)$. Especially, $H$ is called a subgraph of $G$ \emph{induced by} vertices in $V(H)$ when any edge in $G$ whose
endpoints are both in $V(H)$ is an edge in $H$.

We now describe a decoding scheme for a WBAN via graph theory.
Given a WBAN {\em coding scheme} $\cC=\{f_1,...,f_{rn}\}$ (a coding scheme consists of $rn$ functions since 
there are $rn$ packets to be sent to the MS), 
we generate a (multi)graph representation $G=G_{\cC}$ for $\cC$ as follows.
\begin{enumerate} 
\item $G$ has as vertices $1, 2, \ldots, n$.
\item For $i\not =i'$, $(i,i')$ is an edge of $G$ if $P_i\oplus P_{i'}=f_a(P_1,...,P_n)$ for some $f_a$ (\emph{i.e.}, if $P_i\oplus P_{i'}$ is sent to the MS).
\item $G$ has a loop at $i$ if $P_i=f_a(P_1,...,P_n)$ for some $f_a$ (\emph{i.e.}, if $P_i$ is sent to the MS).
\end{enumerate}
The erasure of packets during a transmission can be identified with deletions of edges in $G$. 
Clearly, for a WBAN with $n$ sensors and redundancy $r$, any graph representation of a corresponding coding scheme must have $n$ vertices and $rn$ edges. 

\section{Decodable Graphs\label{decodable}}
We provide a necessary and sufficient condition for full data retrieval at the MS. %\\[-18pt]

\begin{theorem} 
Let $\cC=\{f_1,...,f_{rn}\}$ be a WBAN coding scheme, 
where each $f_a$ is an encoding of packets $P_1,...P_n$,
and $G=G_{\cC}$ be the graph representation for $\cC$.
Now let $H$ be a subgraph of $G$ formed by deleting edges of $G$ corresponding to 
packet erasures occurring in the transmission. 
The MS can retrieve ({\it i.e.}, decode) all packets $P_1,...P_n$ if and only if each connected component in $H$ has at least one loop. 
\end{theorem}
\mbox{}\\[-4ex]
\emph{Proof\/}:
We first show that having a loop in each  component of $H$ 
is sufficient to retrieve all packets $P_1, P_2, \ldots, P_n$. 
Clearly, it is enough to show that for a component with a loop, 
%all packets in the component can be retrieved at the MS.
all packets $P_i$, where $i$ is a vertex in the component, can be retrieved at the MS. 
 
Let $C$ be a connected component with a loop in $H$. A vertex $i$ in $C$ with a loop
signifies that $P_i$ has been received at the MS. Now pick another vertex $j$ in $C$.
Since $i$ and $j$ are in the same connected component, there exists a path $\pi = i,a_1,\cdots,a_{\ell},j$ from $i$ to $j$ 
in $H$, which means that $P_i\oplus P_{a_1}$, $P_{a_m} \oplus P_{a_{m+1}}$ ($1\leq m\leq \ell-1$) and $P_{a_{\ell}}\oplus P_j$ have been received at the MS. 
Then the MS decoder can retrieve $P_j$ from $P_i$ and $P_i\oplus P_j$.

We prove the converse by contradiction. Let $C'$ be a connected component of $H$ with no loops. 
Then, for the $\ell'$ vertices $i_1, i_2, \ldots, i_{\ell'}$ in $C'$, 
the MS can compute only $P_{i_j}\oplus P_{i_s}$, $1\leq j<s\leq \ell'$.
Since these correspond to a system of equations of rank at most $\ell'-1$ over $GF(2)$
the decoder cannot uniquely determine all $P_{i_1}, P_{i_2}, \ldots, P_{i_{\ell'}}$. 
\endproof\mbox{}
 
Therefore, the existence of loops at each component plays an important role in 
selecting a graph for a WBAN coding scheme. For the remainder, we call 
a graph $G$ \emph{decodable} if each of its connected components has a loop, and
denote by $\cD(n,m)$ the set of decodable graphs with $n$ vertices and $m$ edges. 
Otherwise we say that $G$ is called \emph{undecodable}.

Given $G \in \cD(n,m)$, we define 
a \emph{loop cut} to be a subset $\cL$ of $E(G)$
such that $G - \cL$ is undecodable. 
We write $m(G)$ to denote the smallest cardinality of any loop cut of $G$. \\[-18pt]
\begin{remark} For $m(G)$ of a graph $G$, we note the following.
\begin{enumerate}
\item $m(G)\leq \min(L_G, \delta _{I}(G))$ since 
deleting all loops in $G$ or deleting all edges attached to a vertex $\v$ with incidence degree $d_{I}(v)= \delta _{I}(G)$
yields an undecodable graph. 
\item If $L_G\geq \kappa _G$, then $\kappa _G\leq m(G)$
since a resulting graph $\tilde G$ of $G$ after deletion of some edges cannot be undecodable, if $L_{\tilde G}\not =0$, 
unless $\tilde G$ is disconnected. 
\end{enumerate}
\label{minloop_rem}
\end{remark}\mbox{}\\[-20pt]

The robustness of a WBAN coding scheme to packet loss can be measured as a function of 
the number of decodable subgraphs found upon deleting some edges.

\section{The decoding probability of a WBAN \label{decodable_probability}}

Given a graph $G\in \cD(n,m)$, 
we denote by $c_x^G$ the number of decodable subgraphs of $G$ formed by deleting $x$ edges of $G$ and we write $k^G_x = \binom{m}{x}-c_x^G$ to denote the number of undecodable subgraphs of $G$
found by deleting some $x$-set of its edges. 
We define the \emph{decoding probability} of $G$ by 
$$\cP_G:=\sum_{x=0}^{m}c_x^G p^{m-x}q^x,$$
where $p$ is the probability that an edge is not deleted (\emph{i.e.}, the probability that 
a packet is successfully transmitted to the MS)
and $q=1-p$ the probability that an edge is deleted 
(\emph{i.e.}, the probability that a packet is erased during the transmission).
Our interest is to construct a coding scheme $\cC$ for a fixed WBAN whose corresponding 
graph $G=G_{\cC}$ has a high decoding probability.

\begin{lemma}
Let $G\in \cD(n,m)$. Then $m \geq n$. 
\label{edgenumber_lem}
\end{lemma}
\mbox{}\\[-4ex]
\emph{Proof\/}:
It is well known that a connected graph $T$ with $z$ vertices has least $z -1$ edges, with equality if and only if $T$ is a tree. 
Therefore, a connected graph with $z $ vertices is decodable 
only if it has at least $z $ edges since it must contain loops. 
Now let $G$ have connected components $C_i$, $i=1,...,h$. 
Then $|E(C_i)| \geq |V(C_i)|$ for each $i$ and 
hence $m = \sum_{i=1}^{h}|E(C_i)|\geq \sum_{i=1}^{h}|V(C_i)|=n$.
\endproof \mbox{} %\\[-20pt]

From Lemma~\ref{edgenumber_lem}, we immediately deduce that $c_x^G=0$ for 
$x\geq m-n+1$ and so $\cP_G=\sum_{x=0}^{m-n}c_x^Gp^{m-x}q^x$.
Moreover, 
$$\cP_G=\prod_{i=1}^{h} \cP_{C_i},$$
when $G$ consists of $h$ components $C_1,...,C_h$. 

The question of a graph $G$ having optimal decoding probability is related to $m(G)$, which 
is the smallest number $x$ for which $c_x^G <\binom{m}{x}$.
The next lemma gives an upper bound on $m(G)$. \\[-18pt]

\begin{lemma}
Let $G\in \cD(n,m)$. Then $\delta _{I}(G)\leq 2m/n-1$ and $m(G) \leq 2m/(n+1)$.
In particular, $m(G) \leq \min(  \lfloor 2m/n-1 \rfloor , \lfloor 2m/(n+1)  \rfloor)$
\label{mimloopcut_lem}
\end{lemma}
\mbox{}\\[-4ex]
\emph{Proof\/}: 
Since $G$ is decodable, $L_G \geq 1$, and we have
$ n \delta_I(G) \leq S_I(G) = 2m - L_G < 2m.$
Furthermore, since $m(G)\leq L_G$, we have 
$nm(G) \leq n \delta_I(G) \leq S_I(G) = 2m - L_G < 2m -m(G)$. 
\endproof\mbox{}\\[-20pt]
 
It follows that for a WBAN $\cW$ with $n$ packets and redundancy $r$, any graph representation 
$G$ of a coding scheme for $\cW$ satisfies
$m(G)\leq \min(2r-1, \lfloor \frac{2rn}{n+1}\rfloor) = \lfloor \frac{2rn}{n+1}\rfloor$, which is simply 
$2r-1$ whenever $r\leq \frac{n+1}{2}$. 
%As for $M:=\min(2r-1, \frac{2rn}{n+1})$, observe that $M=2r-1$ when $r<\frac{n+1}{2}$ 
%and in general, the inequality holds. So at this moment, we assume $M=2r-1$.
The following proposition shows that it is indeed possible to generate 
some $G$ for which equality in the above holds.
Note that the subscripts $i$ of $P_i$ are computed modulo $n$ in what follows,
if not indicated explicitly. 

\begin{algorithm}[h]
\caption {: A coding scheme for a WBAN $\cW$ with $n$ packets, $k$ relays and redundancy $r$.}
\begin{algorithmic}
\REQUIRE Suppose $L_G=yk+z$, $0\leq z\leq k-1$.
\WHILE {$0\leq j\leq k-1$}
\STATE {let $f_{jt+1}, f_{jt+2}, \ldots, f_{(j+1)t}$ be $t(=nr/k = rs)$ packet encodings of  relay $R_j$.}
\FOR {$t$ packets $P_{js+1}, P_{js+2}, \ldots, P_{js+t}$  received by $R_j$}
\IF  {$b\leq s-1$ } 
	\STATE set $f_{jt+b}:=P_{js+b}\oplus P_{(j+1)s+b}$.
\ELSE
\IF  {$s\leq b\leq t-1$} 
	\STATE set $f_{jt+b}:=P_{js+b}\oplus P_{js+b+1}$. 
\ELSE
\IF  {$b=t$} 
	\STATE $f_{jt+b}:=P_{js+t}\oplus P_{js+1}$. 
\ENDIF
\ENDIF	
\ENDIF	
\IF {$b\leq y$}  
\STATE reset $f_{jt+b}$ to be $f_{jt+b}:=P_{js+b}$.
\ELSE
	\IF {$j+1\leq z$} 
	\STATE reset $f_{js+y+1}$ to be $f_{js+y+1}:=P_{js+y+1}$. 
	\ENDIF
\ENDIF
\ENDFOR
\ENDWHILE
\RETURN {$\cC=\{f_{1}, f_{2}, \ldots, f_{rn}\}$ as a coding scheme.}
\end{algorithmic}
\label{interencoding_algo}
\end{algorithm}

\begin{table}[htb]
\begin{center}
\tiny
\begin{tabular}{|c||c|c|c|c|c|c|}
\hline
Relay & \multicolumn{6}{|c|}{Inter-encoding} \\
\hline
$R_0$ & $P_1$  & $P_2$ & $P_3\oplus P_4$ &  $P_4\oplus P_5$ &  $P_5\oplus P_6$ &  $P_6\oplus P_1$ \\
\hline
$R_1$& $P_4$  & $P_5\oplus P_8 $ & $P_6\oplus P_7$ &  $P_7\oplus P_8$ &  $P_8\oplus P_9$ &  $P_9\oplus P_4$\\
\hline
$R_2$ & $P_7$  & $P_8\oplus P_{11} $ & $P_9\oplus P_{10}$ &  $P_{10}\oplus P_{11}$ & $P_{11}\oplus P_{12}$ & $P_{12}\oplus P_7$\\
\hline
$R_3$& $P_{10}$  &$P_{11}\oplus P_2 $& $P_{12}\oplus P_{1}$ & $P_{1}\oplus P_{2}$ & $P_{2}\oplus P_{3}$ & $P_{3}\oplus P_{10}$\\
\hline
\end{tabular}
\end{center}
\caption{The coding scheme under Algorithm~\ref{interencoding_algo} for 12 sensors, 4 relays and redundancy 2 when $L_G=5$}
\label{example_table}
\end{table}\mbox{}\\[-50pt]

\begin{proposition}
Let $\cC$ be the coding scheme for a WBAN with $n$ packets, $k$ relays and redundancy $r$,
where $k,r \geq 2$,
defined as in Algorithm~\ref{interencoding_algo}. Let $s=n/k$ and let the graph representation $G$ of $\cC$ satisfy
$L_G\geq 2r-1$. If $k\leq L_G\leq (s-1)k$ then it holds that
$m(G)=\delta _{I}(G)=2r-1$.
\label{match_prop}
\end{proposition}
\mbox{}\\[-4ex]
\emph{Proof\/}: 
Observe that the graph $G$ satisfies the following.
\begin{enumerate}
\item Each vertex $i$ with $i \equiv 1 \pmod s$ has a loop.
\item For each $1\leq i \leq n$, the number of edges between vertices $i$ and $i+1$ is $r-1$.
\item $G$ cannot be disconnected without deleting the (multi)edges $(a,a+1)$ and $(a',a'+1)$ for $a\not =a'$.
\item If a vertex $i$ does not have a loop, then
\begin{itemize}
\item it is adjacent to the vertex $i+s$ when $i\not\equiv 0 \pmod s$.
\item it is adjacent to the vertex $i-t+1$ otherwise.
\end{itemize}
\end{enumerate} 
It is straightforward to see that $G$ is connected and 
$\delta _{I}(G)=2r-1$ holds, 
so the edge-connectivity  $\kappa _G$ of $G$ satisfies 
$\kappa _G\leq 2r-1 \leq L_G$. 
Also, we can obtain from Properties 2) and 3) that $\kappa _G\geq 2(r-1)=2r-2$, 
which automatically implies $m(G)\geq 2r-2$ from 2) in Remark~\ref{minloop_rem}.

Now suppose that edges described in Property 3) are deleted from $G$,  
and call the resulting graph $\hat G$.
Denote by $H_a$ the subgraph of $\hat G$ induced by the vertices $a+1,a+2,...,a'$, %$\pmod n$, 
and by $H_{a'}$ the one induced by the vertices $a'+1,a'+2,...,a$. % $\pmod n$. 
%Then, if we see $H_a$ and $H_{a'}$ within $G$,  either there exists some edges 
%between $H_a$ and $H_{a'}$ or not. 
If both $H_a$ and $H_{a'}$ contain loops, then we can conclude that 
$m(G)\geq 2r-1$. 
Furthermore, it cannot happen that neither $H_a$ nor $H_{a'}$ have loops since 
$L_G=L_{\hat G}\geq 1$. Therefore, we need only to consider the case (without loss of generality)
when  $H_a$ contains a loop but $H_{a'}$ does not. In this case, 
we will see $H_a$ and $H_{a'}$ within $\hat G$ and show the existence of an edge in $E(\hat G)$ joining them, which implies that $m(G)\geq 2r-1$. 

As $H_{a'}$ does not contain loops, $|V(H_{a'})|<s$ since otherwise,  
at least one of the vertices $i$ in $H_{a'}$ satisfies 
$i \equiv 1 \pmod s$, 
and therefore, $H_{a'}$ contains a loop from Property 1).

If $H_{a'}$ contains a vertex $i$ with $i\equiv 0 \pmod s$, 
$i$ is adjacent to the vertex $i-t+1$, where $i-t+1 \equiv 1 \pmod s$ as $t=sr$. 
Since $i-t+1$ has a loop from Property 1), it is in $H_a$. 
If each vertex $i$ in $H_{a'}$ satisfies $i\not\equiv 0 \pmod s$, then $i$ is adjacent to 
$i+s$. Since $|V(H_{a'})|<s$ and $|V(H_{a})|>n-s=(k-1)s\geq s$, $i+s$ is in $H_a$.
In each case, there exists an edge in $E(\hat G)$ joining $H_a$ and $H_{a'}$ as required. 
\endproof\mbox{}\\[-20pt]

%%%%%%%%%%%%%%%%%%%%%%%%%%%%%%%%%%%%%%%%%%%%%%%%%%%%%%%%%%%%%%%%%%%%%%%%%%%%%%%%%%%%%%%%%%%%%%%

We now present some upper bounds on $c_x^G$ for $x$.
The following lemma is a sharp upper bound on $c_x^G$ when $x=m(G)$. \\[-18pt]

\begin{lemma}\label{lemub1}
Let $G$ be a decodable graph with $n$ vertices and $m$ edges.
Then $c^G_{m(G)}\leq \binom{m}{m(G)} - m(G)(n+1)-n+2m-1$ 
\end{lemma}
\mbox{}\\[-4ex]
\emph{Proof\/}: 
First recall that $m(G) \leq \min(L_G,\delta_I(G))$.
Let $\alpha $ be the number of vertices of $G$ with incidence degree $m(G)$ and let
$\beta$ be the number of vertices with incidence degree at least $m(G)+2$. 
Then 
$$m(G)\alpha+(m(G)+1)(n-\alpha-\beta)\leq 2m-L_G-(m(G)+2)\beta,$$ which implies that
$$\alpha \geq L_G+ nm(G)+n -2m + \beta \geq m(G)(n+1) + n -2m,$$ since 
$L_G \geq m(G)$ and $\beta \geq 0$.
Clearly $c_{m(G)}^G \leq \binom{m}{m(G)} - \alpha $, since deleting any $m(G)$ edges incident with a vertex
of incidence degree $m(G)$ results in an undecodable graph.
Since $G$ is decodable, no vertex of incidence degree $m(G)$ is incident with all loops of $G$. 
Therefore, if $\alpha=m(G)(n+1) + n -2m$, then $L_G=m(G)$ and so $c_{m(G)}^G \leq \binom{m}{m(G)} - \alpha -1$ 
since we also have to count the case of deleting all $m(G)$ loops from $G$. If $\alpha > m(G)(n+1) + n -2m$, the result follows trivially.
\endproof\mbox{}\\[-20pt]

Clearly, $k^G_{m(G)} \geq m(G)(n+1)-n+2m-1$ for any $G \in {\cD(n,m)}$.
We can also show a tight upper bound when $x$ is close to $m(G)$ .\\[-18pt]

\begin{lemma}\label{lemub1eq}
Let $G\in \cD(n,m)$ satisfy
$k^G_{m(G)} = m(G)(n+1)+n-2m+1$. Then, with the same notation as in Lemma \ref{lemub1}, $\beta = 0$ and either
\begin{enumerate}
   \item
       $\alpha = m(G)(n+1)+n-2m$ and $L_G=m(G)$, or
   \item
       $\alpha =m(G)(n+1) + n - 2m +1$ and $L_G = m(G)+1$.    
\end{enumerate} 
\end{lemma}
\mbox{}\\[-4ex]
\emph{Proof\/}: 
Let $\theta = m(G)(n+1)+n-2m$. Recall that, as in the proof of Lemma \ref{lemub1},
\begin{equation}\label{eqalpha}
\alpha \geq L_G+ nm(G)+n -2m + \beta \geq m(G)(n+1) + n -2m =\theta.
\end{equation}
Therefore, 
$$ \theta + 1 = k^G_{m(G)} \geq \alpha \geq \theta,$$
so either $\alpha = \theta$, in which case $k^G_{m(G)}= \alpha+1$,
or $\alpha = \theta + 1$ and $k^G_{m(G)}=\alpha$.  
For the case $\alpha = \theta$ from (\ref{eqalpha}), we must have $L_G=m(G)$ and $\beta = 0$. 
In the latter case we have
$\alpha = \theta+1 \geq \theta - m(G)+L_G + \beta,$ which gives 
$m(G)+1 \geq L_G + \beta \geq m(G)+ \beta$.
Therefore, either $\beta = 1$ and $L_G=m(G)$ or $\beta = 0$ and $L_G=m(G)+1$.
Since for $\alpha=\theta+1$ we have $k^G_{m(G)}=\alpha$, every undecodable subgraph of $G$ found by deleting $m(G)$ edges is constructed by deleting the $m(G)$ edges that meet a vertex of incidence degree $m(G)$.
If $L_G=m(G)$ then $G$ has a vertex of incidence degree $m(G)$ that is incident with every loop of $G$, contradicting the decodability of $G$, so we deduce that $\beta = 0$ and $L_G=m(G)+1$.
\endproof\mbox{}\\[-20pt]

\begin{lemma}\label{lemub2}
Let $G$ be a graph with $m>2$ and satisfying the hypothesis of Lemma \ref{lemub1eq}.
Let $\theta = m(G)(n+1)+n-2m$.
Then $\Delta_\ell(G) \leq m(G)-1$ and
$$k_{m(G)+x}^G\geq (\theta + 1)\binom{m-m(G)}{x} + (n-\theta)\binom{m-m(G)-1}{x-1},$$
for any $x$ satisfying $1 \leq x \leq m(G)- \Delta_\ell(G)$.
\end{lemma}
\mbox{}\\[-4ex]
\emph{Proof\/}: 
Clearly, since $G$ is decodable, $\delta_\ell(v)\leq m(G)-1$ for any $v\in V(G)$ such that $\delta_I(v)=m(G)$, and $\delta_\ell(v)\leq m(G)$ for any $v\in V(G)$ satisfying $\delta_I(v)=m(G)+1$. 

Suppose that $\alpha = \theta$ (\emph{i.e.}, that 1) of Lemma~\ref{lemub1eq} holds). Then 
$L_G=m(G)$ and $k_{m(G)}^G=\alpha+1$.
Suppose that $v \in V(G)$ has incidence degree $m(G)+1$. 
If $\delta_{\ell}(v) = m(G)$ then $v$ has exactly one neighbour,
so an undecodable subgraph results by deleting the only non-loop edge incident with $v$ and we deduce that $m(G)=1$.
Then $0 \leq \alpha = 2(n-m)+1 \leq 1$, since $n \leq m$, which forces $n=m$ and $\alpha = 1$.
It follows that $G$ is a path graph with a single loop at a vertex of incidence degree 2 and one leaf (a vertex of degree $1$). Therefore, $k^G_{m(G)} = k^G_1=m=n$. On the other hand, $k^G_1 = \alpha+1=2$, contradicting our assumption that $m>2.$
We deduce that $\Delta_\ell(G) \leq m(G)-1$. 

For the case $\alpha = \theta+1$ (\emph{i.e.}, if 2) of Lemma~\ref{lemub1eq} holds), we have
$L_G=m(G)+1$ and $k^G_{m(G)}=\alpha$. If $v$ is a vertex of $G$ with $\delta_I(v)=m(G)+1$ and 
$\delta_{\ell}(v)=m(G)$, then an undecodable subgraph with $m-2$ edges results by deleting the only non-loop edge of $v$ and the single loop not incident with $v$. Then $m(G) \leq 2.$
If $m(G)=2$ then $\alpha = k^G_{m(G)}=k^G_2 \geq \alpha+1$. 
If $m(G)=1$ then $0\leq \alpha =2(n-m)+2$ so that either $n=m$ and $\alpha=2$ or $n=m-1$ and
$\alpha=0.$ In the former case, $G$ must have exactly two connected components, each of which is a path graph with exactly one loop at a vertex of incidence degree 2 and one leaf. Then $2=\alpha = k^G_{m(G)}=m=n$, giving a contradiction to $m>2$. In the latter case, $G$ is a path graph with exactly 2 loops and no leaves, so deleting a single edge never results in an undecodable subgraph, contradicting $m(G)=1$. 
It follows that $\Delta_{\ell}(G) \leq m(G)-1.$

Let $x \in \{1,..., m(G)- \Delta_\ell(G)\}$. Consider the following operations, each of which results in an undecodable subgraph of $G$ with $m-m(G)-x$ edges.
\begin{enumerate}
   \item
      Delete $m(G)$ edges incident with a vertex of incidence degree $m(G)$ and delete a further 
      $x$ edges arbitrarily.      
   \item
      Delete $m(G)+1$ edges incident with a vertex of incidence degree $m(G)+1$ and delete a further $x-1$ edges arbitrarily. 
   \item
      Delete all $L_G$ loops of $G$, and then delete a further $m(G)+x-L_G$ edges arbitrarily.   
\end{enumerate}
Observe first that no two distinct vertices of incidence degree $d$ are coincident with $d$ edges, since $G$ is decodable, so there are exactly $\alpha\binom{m-m(G)}{x}$ (respectively $(n-\alpha)\binom{m-m(G)-1}{x-1}$) ways to produce an undecodable subgraph by the operation 1) (respectively, by the operation 2)). The operations 1) and 2) are mutually exclusive, since in 1) at most $x \leq m(G)-1$ edges are deleted from a vertex of incidence degree $m(G)+1$. Moreover, the operations
2) and 3) are exclusive to each other, since in 2) at most 
$$\delta_\ell(v) + x - 1 \leq m(G) - (\Delta_\ell(G)-\delta_\ell(v)) - 1 \leq m(G) - 1$$ loops are deleted, for any vertex $v$ of incidence degree $m(G)+1$.

For the case $\alpha=\theta+1$, 1) and 3) are exclusive, since $m(G)<L_G$, and at most $\delta_\ell(v)+x \leq m(G) - (\Delta_\ell(G)-\delta_\ell(v))\leq m(G)$ loops are deleted for any given vertex $v$ of degree $m(G)$. 

Now suppose that $\alpha=\theta$ and let $v\in V(G)$ such that $\delta_I(v)=m(G)$ and $\delta_\ell(v) \geq 1$. The following actions result in an undecodable subgraph by deleting some $m(G)$ edges of $G$.   
\begin{enumerate}
   \item[(a)]
      Delete $m(G)$ edges incident with a vertex of incidence degree $m(G)$.
   \item[(b)]
      Delete all $m(G)$ loops.
   \item[(c)]
      Delete the $m(G)-\delta_\ell(v)$ non-loops edges incident with $v$ and delete the remaining $L_G - \delta_\ell(v)$ 
      loops of $G$ that are not incident with $v$.   
\end{enumerate}
Clearly under the assumption $\delta_\ell(v) \geq 1$, the operations (a),(b) and (c) are pairwise exclusive and so $\alpha+1 = k^G_{m(G)} \geq \alpha + 2$, giving a contradiction, so we deduce that no vertex of incidence degree $m(G)$ is incident with a loop. Then in 1), for a given vertex $v$ satisfying $\delta_I(v) = m(G)$, at most $\delta_\ell(v) + x =x \leq m(G)-1$ loops are deleted, which means 1) and 3) are mutually exclusive.

It follows that 
\begin{eqnarray*}
k_{m(G)+x}^G & \geq & \alpha \binom{m-m(G)}{x}+(n-\alpha)\binom{m-m(G)-1}{x-1} \\
             &  +   & \binom{m-L_G}{m(G)+x-L_G}, 
\end{eqnarray*}
 which yields
$$k_{m(G)+x}^G\geq (\theta + 1)\binom{m-m(G)}{x} + (n-\theta)\binom{m-m(G)-1}{x-1},$$
for any $G \in \cD(n,m)$.
\endproof\mbox{}\\[-20pt]

For given $c^G_{x}$ we can compute 
an upper bound on $c^G_{x+z}$ for $z\geq 0$ by using the following easy result.\\[-18pt] 

\begin{lemma}\label{ub_lem}
   Let $G$ be a graph with $n$ vertices and $m$ edges.
   Then
   $$k^G_{x+z} \geq k^G_{x} \binom{m-x}{z} / \binom{x+z}{z}$$ 
   for $z \geq 0$. 
\end{lemma}
\mbox{}\\[-4ex]

The following corollary is now immediate.\\[-18pt]

\begin{corollary}
Let $G \in \cD(n,m)$ satisfy the hypothesis of Lemma \ref{lemub1eq}.
Then for each $z \geq 0$
\begin{eqnarray*}
k^G_{2m(G)-\Delta_\ell(G)+z}  \geq 
k^G_{2m(G)-\Delta_\ell(G)} \frac{\binom{m-2m(G) + \Delta_\ell(G)}{z}} { \binom{2m(G)-\Delta_\ell(G)+z}{z}}.
\end{eqnarray*} 
\end{corollary}
%\mbox{}\\[-4ex]

%\begin{corollary}\label{ub_2r-1_lem}
%   Let $G \in {\cC}(n,rn)$ be a graph representation for a WBAN coding scheme %such that $m(G)=2r-1$.
%   Then $c^G_{2r-1} \leq \binom{rn}{2r-1}-2r$. Moreover, if equality holds then
%   $c^G_{2r+z} \leq \binom{rn}{2r+z}-(2rn^2-4r^2+n+1)\binom{rn-2r}{z} / %\binom{2r+z}{z}$
%   for each $z \geq 0.$
%\end{corollary}

\section{Simulation Results of WBANs \label{simulation}}
In this section, we will provide simulation results and 
see the correspondence between simulation results and theoretical results that have been discussed in this paper. 
We focus on coding schemes of WBANs with 9 sensors (which implies 9 packets), 3 relays and redundancy 2. 
More precisely, we follow the coding scheme introduced in 
Algorithm~\ref{interencoding_algo} for $1\leq L_G\leq 9$ and the one with no inter-encoded packets, as presented in Table~\ref{interencoding_table}.

\begin{table}[h]
\begin{center}
\footnotesize
\begin{tabular}{|c||c|c|c|c|c|c|}
\hline
Relay & \multicolumn{6}{|c|}{Inter-encoding} \\
\hline
$R_0$ & $P_1$  & $P_2 $ & $P_3$ &  $P_4$ &  $P_5 $ &  $P_6 $ \\
\hline
$R_1$& $P_4$  & $P_5 $ & $P_6$ &  $P_7$ &  $P_8$ &  $P_9$\\
\hline
$R_2$ & $P_7$  & $P_8$ & $P_9$ &  $P_{1}$ & $P_{2} $ & $P_{3} $\\
\hline
\end{tabular}
\end{center}
\caption{No inter-encoded packets}
\label{interencoding_table}
\end{table}
%\mbox{}\\[-20pt]

Now let $G_i$ and $G$ the graph representations of the coding scheme with $L_G=i$ and the one with no inter-encoded packets, respectively. 
Since each representation consists of $9$ vertices and $18$ edges, we have from Lemma~\ref{edgenumber_lem} that 
$c_{x}^{G_i}=c_{x}^{G}=0$ for any $i$ whenever $x\geq 10$ . The detailed information on $c_{x}^{G_i}$ and $c_{x}^{G}$ 
for $1\leq x\leq 9$ is given in Table~\ref{number_table}.  The table also contains 
the information on $D_x$'s, which are the upper bounds of  $c_{x}^{H}$'s for $H\in \cD(9,18)$ obtained from Lemmas~\ref{lemub1}, ~\ref{lemub2} and \ref{ub_lem}.

For $3\leq i\leq 9$, observe that $\binom{18}{x}=c_{x}^{G_i}$ for $x=1,2$, which implies  $m(G_i)=3(=2r-1)$.
Furthermore, $c_{x}^{G_3}$ is the largest one amongst all examples for any $1\leq x\leq 9$. 
In addition, $c_{3}^{G_3}$ and $c_{4}^{G_3}$ meet
the upper bounds obtained from Lemmas~\ref{lemub1} and ~\ref{lemub2}.

\begin{table}[h]
\begin{center}
\tiny
\begin{tabular}{|c||c|c|c|c|c|c|c|c|c|}
\hline
 $x$            & $1$ & $2$ & $3$  & $4$  & $5$  & $6$  &  $7$  &  $8$ & $9$   \\
 \hline
$\binom{18}{x}$ & 18 & 153 & 816  & 3060 & 8568 &   18564 & 31824   & 43758 & 48620 \\
 \hline
$D_x$          & 18 & 153 & 812  & 2994 & 8064 &   17472 &  29952   & 41184 & 45760 \\
\hline
$c_{x}^{G_1}$ & 17 & 136 & 677  & 2333 & 5842 &   10803 &  14540 & 13297 & 10340 \\
\hline
$c_{x}^{G_2}$ & 18 & 152 & 797  & 2889 & 7603 &  14769 &  20880 &  20073 & 12365 \\
\hline
$c_{x}^{G_3}$ & 18 & 153 & 812  & 2994 & 8052 &  16053 &  23388 &  23277 & 12500 \\
\hline
$c_{x}^{G_4}$ & 18 & 153 & 812  & 2993 & 8042 &  16008 &  23273 &  23101 & 12365 \\
\hline
$c_{x}^{G_5}$ & 18 & 153 & 811  & 2979 & 7952 &  15660 & 22402 &  21731 & 11273 \\
\hline
$c_{x}^{G_6}$ & 18 & 153 & 810  & 2964 & 7851 &  15260 &  21405 &  20232 & 10192 \\
\hline
$c_{x}^{G_7}$ & 18 & 153 & 809  & 2948 & 7736 &  14779 &  20135 &  18161 & 8532 \\
\hline
$c_{x}^{G_8}$ & 18 & 153 & 808  & 2932 & 7621 &  14299 &  18886 &  16199 & 7053 \\
 \hline
$c_{x}^{G_9}$ & 18 & 153 & 807  & 2916 & 7506 &  13821 &  17667 &  14373 & 5776 \\
\hline
$c_{x}^{G}$   & 18 & 144 & 672  & 2016 & 4032 &  5376 &  4608 &  2034 & 512 \\
\hline
\end{tabular}
\end{center}
\caption{\footnotesize{The number of decodable graphs}}
\label{number_table}
\end{table}
%\mbox{}\\[-30pt]

Using the information, we can derive the decoding probabilities $\cP_{G_i}$ and $\cP_{G}$. 
We provide the decoding probabilities, together with the probabilities 
obtained from simulations in Table~\ref{dp_table}. 
As for the simulation results, we computed the probabilities $P$ as
$$P=\frac{\mbox{the number of success simulations}}{\mbox{the total number of simulations}},$$
where success simulations mean the ones in which all packets are retrieved.
We ran the programme by setting the total number of simulations to be 5000000.
We can see that applying coding scheme increases the 
decoding probability remarkably.

%We can see that decoding probabilities 
%are close enough to the ones obtained from simulations. 

\begin{table}[h]
\begin{center}
\scriptsize
\begin{tabular}{|c||c|c|}                 
\hline
            & Decoding probability & Simulation results \\
\hline            
$G_1$ &$\cP_{G_1}=0.7728010935$ & 0.77262 \\
\hline
$G_2$ & $\cP_{G_2}=0.9257409618$ &0.92564 \\
\hline
$G_3$ &$\cP_{G_3}=0.9558104057$ &0.95578 \\
\hline
$G_4$ & $\cP_{G_4}=0.9551821038$ &0.95518 \\
\hline
$G_5$ &$\cP_{G_5}=0.9493923505$ & 0.94944 \\
\hline
$G_6$ & $\cP_{G_6}=0.9429367740$ &0.94272 \\
\hline
$G_7$ &$\cP_{G_7}=0.9353111111$ & 0.93524 \\
\hline
$G_8$ & $\cP_{G_8}=0.9277553360$ &0.92766 \\
\hline
$G_9$ &$\cP_{G_9}=0.9202926069$ & 0.92018 \\
\hline
$G$ &$\cP_{G}=0.6924597789$ & 0.69254 \\
\hline
\end{tabular}
\end{center}
\caption{The decoding probabilities with $p=0.8$}
\label{dp_table}
\end{table}
%\mbox{}\\[-20pt]

For any probability $p$, $\cP_{G_3}$ has been the optimal (in terms of decoding probability) amongst 
all graphs in $\cC(9,18)$ at this moment. Indeed, if there exists  $H\in \cC(9,18)$ such that $c_{x}^{H}=D_x$ 
we have $\cP_H-\cP_{G_3}=0.02087697704$ (\emph{resp.} 0.0007125786313) 
when $p=0.8$ (\emph{resp.} $p=0.9$). 

%Since the difference is quite small, 
%we can say that $G_3$ would be a good candidate with the highest decoding probability amongst all graphs in $\cC(9,18)$.

%We have been working to show that $\cP_{G_3}$ is indeed optimal, using results from graph theory such as
%Tutte polynomials. We are also interested in applying the techniques to more general cases. 

%Amongst these WBANs, the ones with 3 unencoded packets and  with 4 uncoded packets 
%provide high decoding probabilities. This matches the 
%theoretical results discussed in previous section. Indeed, the 
%decoding probabilities obtained from simulations are close enough 
%to the ones computed mathematically. 

\end{document}